\documentclass[twocolumn,showpacs,preprintnumbers,amsmath,amssymb]{revtex4}

\usepackage{graphicx}
\usepackage{dcolumn}
\usepackage{amsfonts,amsmath,amssymb,bm}
\usepackage{epsfig}

\begin{document}

\title{\bf {$I$-$V$ curves of Fe/MgO (001) single- and double-barrier tunnel junctions}} 
\date{\today} 

\author{J. Peralta-Ramos$^a$}
\email[Corresponding author: ]{peralta@cnea.gov.ar}
\author{A. M. Llois$^{a,b}$}
\author{I. Rungger$^c$}
\author{S. Sanvito$^c$}

\affiliation{a)Departamento de F\'isica, Centro At\'omico Constituyentes, Comisi\'on 
Nacional de Energ\'ia At\'omica, Buenos Aires, Argentina\\
b) Departamento de F\'isica, Facultad de Ciencias Exactas y Naturales,
Universidad de Buenos Aires, Buenos Aires, Argentina\\
c) School of Physics and CRANN, Trinity College, Dublin 2, Ireland}

\begin{abstract}

In this work, we calculate with {\it ab initio} methods the current-voltage characteristics for 
ideal single- and double-barrier Fe/MgO (001) magnetic tunnel junctions. 
The current is calculated in the phase-coherent limit by using the recently developed {\it Smeagol}
code combining the non-equilibrium Green's functions formalism with density functional theory. 
In general we find that double-barrier junctions display a larger magnetoresistance, which decays
with bias at a slower pace than their single-barrier counterparts. This is explained in terms of 
enhanced spin-filtering from the middle Fe layer sandwiched in between the two MgO barriers. 
In addition, for double-barrier tunnel junctions 
we find a well defined peak in the magnetoresistance at a voltage of $V=0.1$~Volt. This is
the signature of resonant tunneling across a majority quantum well state. Our findings are
discussed in relation to recent experiments.

\end{abstract}

\maketitle

\section{Introduction}

Tunneling magnetoresistance (TMR) arises in magnetic tunnel junctions consisting of an insulating barrier 
sandwiched by two ferromagnetic electrodes. It is observed that the resistance strongly depends on the relative 
orientation of the magnetization vectors of the electrodes. It is usually small when the magnetizations 
are parallel to each other and increases when they are antiparallel, and the change in resistance can be large. The TMR 
coefficient is defined as 
$TMR=[(I_\mathrm{P}-I_\mathrm{AP})/I_\mathrm{AP}]\times100$, where $I_\mathrm{P}$ and $I_\mathrm{AP}$
are the currents in the parallel (P) and in the antiparallel (AP) magnetic configuration related to the same applied voltage, respectively. Since the 
pioneering work of Julli\`ere \cite{julliere,fabian}, the experimentally attainable TMR values have been steadily increasing, mainly due to the tremendous advances in growth techniques. In particular a steep increase in TMR magnitude followed from the growth of highly crystalline Fe/MgO (001) tunnel junctions with atomically flat surfaces 
\cite{yuaseJPD}. Currently the record value is 500 $\%$ at room temperature in fully epitaxial FeCoB/MgO/FeCoB
junctions \cite{Ohno1}. This value is much larger than the one it can be obtained in standard metallic giant 
magnetoresistive devices, and this explains why magnetic tunnel junctions will soon replace  
metallic spin-valves in the read heads of near-future hard disk drives \cite{fuchs}.

In epitaxial single-barrier magnetic tunnel junctions (SBMTJs) the large TMR is a consequence of the symmetry 
matching between Bloch states in the electrodes and the complex bands within the insulator \cite{mav}. These
latter are evanescent states lying in the band gap and decaying inside the spacer. For thin spacers, another 
transport mechanism may arise, when surface states at each side of the barrier come into resonance. These 
form the so-called {\it hot spots} in the transmission coefficient as a function of the transverse wave-vector 
$k_\parallel$.  This mechanism was clearly demonstrated in the {\it ab initio} calculations of MacLaren {\it et al.} 
\cite{mc} for Fe/MgO SBMTJs, in which a Fe minority surface state was found resonant through the MgO spacer.
It was then shown experimentally by Tiusan {\it et al.} \cite{tiusan04} and theoretically by Rungger {\it et al.} 
\cite{rungger} that, in Fe/MgO SBMTJs, these surface states are very sensitive to the applied bias voltage $V$, and 
as a consequence the TMR is maximal at zero bias and significantly decreases with increasing bias. The bias voltage 
at which the TMR drops to half of its zero bias value is denoted as $V_{1/2}$ and it is taken as one of the 
quality factors for the applicability of SBMTJs in real devices. Typically $V_{1/2}$ is less than 0.7~Volt.

Nowadays, one of the main challenges in the field of spin-electronics is to reach simultaneously large TMR and 
$V_{1/2}$ values. One possible route to accomplish this goal was proposed by Zhang \cite{zhangcl} a decade 
ago and consists in the insertion of a metallic slab in between the insulating spacer, thus to form a double-barrier 
magnetic tunnel junction (DBMTJ). In this architecture, quantum well states (QWSs) formed by confinement
are found in the in-between metallic slab. These can come into resonance with the evanescent states inside the 
barriers, enhancing the TMR in a significant way and allowing the tuning of the TMR by controlling the thickness of
the in-between slab. This tunability of the TMR arises because changing the thickness of the in-between metallic slab produces a shift in the energy of the QWS, and 
creates new ones. In addition, since the potential drop is shared across the two barriers and 
an additional spin-dependent scattering potential is introduced by the in-between magnetic slab, the $V_{1/2}$ for
DBMTJs is expected to be larger than that of a SBMTJ of same combined thickness. 

To date there have been several theoretical studies on the spin-dependent properties of DBMTJs \cite{pet,wil,wang,nos}, with 
either magnetic or non-magnetic in-between slabs. Usually these are conducted at zero bias in the linear response limit or in 
a non-self-consistent fashion. The key features emerging from these studies are, first that the TMR can reach extremely large values under resonant conditions, and secondly that the TMR can be enhanced not only by resonant tunneling through QWS 
but also by the {\it spin-filter effect} (SFE) \cite{nos}. This effect is a consequence of the insertion of a {\it magnetic} 
slab in between the barriers. In fact in these DBMTJs the P configuration corresponds to all the magnetizations 
(the two electrodes and the in-between layer) being aligned parallel to each other, while in the AP configuration
the electrode magnetizations remain parallel to each other but the in-between magnetization is antiparallel to them. 
Thus in the AP configuration of DBMTJs there are effectively two interfaces where the magnetization vector
change sign and this greatly enhances backscattering. Since the electrodes and the in-between Fe layers have 
different coercive fields, due to their different thicknesses, these magnetic configurations 
P and AP are experimentally attainable \cite{fuchs}. 

The SFE has been clearly shown in our previous calculations for ideal Fe/ZnSe (001) 
SBMTJs and DBMTJs with Fe in-between layers \cite{nos}. The main difference between resonant tunneling through 
QWS and the SFE is that the former strongly depends on the thickness of the in-between slab, while the latter is almost 
independent. Thus, DBMTJs show a variety of spin-dependent transport phenomena richer than that of their
single-barrier counterparts, and in principle they promise a better control of the TMR. Some of these expectations are now confirmed 
experimentally. For example, Nozaki {\it et al.} \cite{noz05} have recently measured the $I$-$V$ curves of fully epitaxial 
Fe/MgO SBMTJs and DBMTJs (with Fe as the in-between metallic slab), and found that (i) the TMR of DBMTJs is 
larger than that of SBMTJs, and (ii) the TMR decrease with bias is {\it significantly} slower in DBMTJs than in SBMTJs. 
Similarly Zeng {\it et al.} \cite{zeng} have shown that CoFeB/Al-O DBMTJs also have TMR and $V_{1/2}$ 
values larger  than those of the corresponding SBMTJs. Other works along the same lines which confirm these results include 
that of reference \cite{iovan}.

In spite of the amount of research carried out on the spin-dependent transport properties of DBMTJs, self-consistent 
calculations of $I$-$V$ curves for a realistic junction from first principles are still lacking. The purpose of this paper 
is to fill this gap. We present {\it ab initio} calculated $I$-$V$ characteristics for Fe/MgO (001) DBMTJs, 
and compare them to those of SBMTJs, relating the transport properties of the devices to their electronic structure. 
To this goal, we use the recently developed {\it ab initio} code {\it Smeagol} \cite{smeagol}, that combines the 
pseudopotential density functional code {\it Siesta} \cite{siesta} for the electronic structure calculations with the non-equilibrium 
Green's function formalism (NEGF) for phase-coherent transport \cite{datta}. 

\section{Calculation details}

\begin{figure}[htb]
\epsfxsize=14cm
\centerline{\epsffile{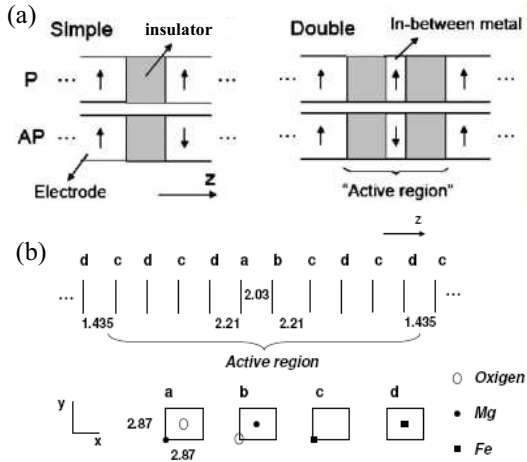}}
\caption{(a) Structure of single- and double-barrier junctions. The arrows indicate the two magnetic configurations considered (parallel/antiparallel). 
(b) Structure of a SBMTJ with $n$=2 MLs, showing the {\it active region} and part of the semi-infinite 
Fe electrodes. The junction is periodic in the $x$-$y$ plane while $z$ is the 
transport direction. Distances are in \AA.}
\label{Fig1}
\end{figure}

Our SBTJs consist of $n$ monolayers (MLs) of MgO (001) sandwiched by two {\it semi-infinite} bcc Fe (001) 
electrodes, while our DBMTJs are multilayers of the type (MgO)$_n$/Fe$_m$/(MgO)$_n$ (001) sandwiched by 
the same electrodes. In both cases, the junctions are assumed to be periodic in the {\it x-y} plane, being $z$ the transport direction. In order to account for the charge transfer and to correctly reproduce the band offset between Fe and MgO, 
we include in the cell for self-consistent calculations four Fe MLs belonging to the electrodes at both sides of the junction.
This is\index{} enough to correctly account for charge screening inside the ferromagnet. Similar to previous calculations 
\cite{wang,rungger,mc} the lattice constant of the electrodes is fixed to 2.87~\AA\ and that of MgO is 
taken to be $\sqrt{2}$ larger. This, together with a 45$^\circ$ rotation of the Fe unit cell, allows  
epitaxial matching between Fe and MgO. Figure~\ref{Fig1} shows the structure of a single-barrier 
junction with $n$=2 MLs (4~\AA), together with the schematic structure of single- and double-barrier junctions. In this work, the possible appearance of FeO interfacial layers, as well as atomic relaxation 
and disorder, are not considered, so that our calculations are valid only in the ballistic limit for atomically ordered structures.  

For the electronic structure of the junctions, we use norm-conserving pseudopotentials, double-zeta basis set 
for all the angular momenta and the generalized gradient approximation (GGA) \cite{PBE} to the exchange and correlation potential. We have thoroughly checked that the band structure and the density of states of 
bulk Fe, bulk MgO and Fe/MgO multilayers,  as well as the charge transfer and magnetic moments in the  
last case, are very well reproduced as compared to FP-LAPW results obtained using the highly accurate 
{\it WIEN2k} code \cite{wien}. We obtain a band offset (the difference between the Fermi energy 
$E_\mathrm{F}$ of Fe and the valence band of MgO) of 3.51 eV, in very good agreement with previous 
theoretical \cite{yu} and experimental reports \cite{klaua}. 
As well-known, density functional calculations using semi-local exchange and correlation functionals
underestimate the band gap and ours are not an exception. We obtain a band gap of 5.4~eV (as compared to the experimental value of 7.8~eV \cite{klaua,whited}), which agrees well with what expected from GGA
\cite{mc,yu}.

The ballistic current density at each bias voltage $V$ is calculated as 
\begin{equation}
I^\sigma (V)=\frac{e}{h}\int dE ~T^\sigma (E,V)[f_\mathrm{L}-f_\mathrm{R}]
\label{current}
\end{equation}
where $f_\mathrm{L}=f(E-\mu_\mathrm{L})$ ($f_\mathrm{R}=f(E-\mu_\mathrm{R})$) is the Fermi-Dirac function 
evaluated at $E-\mu_\mathrm{L}$ ($E-\mu_\mathrm{R}$) with $\mu_\mathrm{L}=E_\mathrm{F}+eV/2$
($\mu_\mathrm{R}=E_\mathrm{F}-eV/2$) the chemical potential of the left (right) electrode. 
Finally $\sigma$ is the spin index standing for majority ($\sigma=\uparrow$) and minority 
($\sigma=\downarrow$) spins. The transmission coefficient $T^\sigma (E,V)$ is calculated for each bias and 
it is given by
\begin{equation}
T^\sigma (E,V)=\frac{1}{V_\mathrm{BZ}}\int dk_x dk_y ~T^\sigma (E,V,k_x,k_y)
\label{k}
\end{equation}
where $V_\mathrm{BZ}$ is the area of the 2D Brillouin zone orthogonal to the transport direction. 
Here we assume that both spin and transverse momentum are conserved, an approximation that is
valid for relatively thin epitaxial junctions. The {\it $k_\parallel$-resolved} transmission coefficient appearing in Eq. (\ref{k}) 
($\vec{k}_\parallel=k_x \hat{x}+k_y \hat{y}$, see Fig.~\ref{Fig1}) is calculated 
from the non-equilibrium Green's functions formalism in the usual way \cite{smeagol,datta}. 
It is given by $T=Tr[\Gamma_\mathrm{L}G^r\Gamma_\mathrm{R}G^a]$, where for simplicity we omit the 
spin label $\sigma$. Here, $\Gamma_\mathrm{L,R}$ are the broadening matrices describing the interaction
(thus the finite lifetime) of the scattering region energy levels with the left- and right-hand side electrodes, 
and $G^{r}$ ($G^a$) is the associated retarded (advanced) Green's function describing the one-electron 
electronic structure of the scattering region. 
The broadening matrices are calculated from the self-energies 
$\Sigma_{L,R}$ as $\Gamma_{L,R}=i(\Sigma_{L,R}-\Sigma^\dag_{L,R})$. These in turn are obtained with 
the semi-analytic method described in reference \cite{san,ivan}. 

In our calculations, we use a $8\times 8\times 8$ k-point mesh in reciprocal space to calculate the density matrix 
of the scattering region and a $150\times 150\times 1$ mesh to evaluate the current at each bias voltage. 
We have carefully verified that these meshes are sufficient for converging the density matrix and the current. 

\section{Results: Single-barrier tunnel junctions}

In order to benchmark our calculations for DBMTJs, we have
calculated first the conductance and TMR at zero and finite bias
of several single-barrier Fe/MgO (001) junctions as a function of
the barrier thickness n. In general we find a rapid decrease of the
TMR as a function of bias in agreement with previous theoretical 
results \cite{rungger}. 
\begin{figure}[htb]
\epsfxsize=7.0cm
\centerline{\epsffile{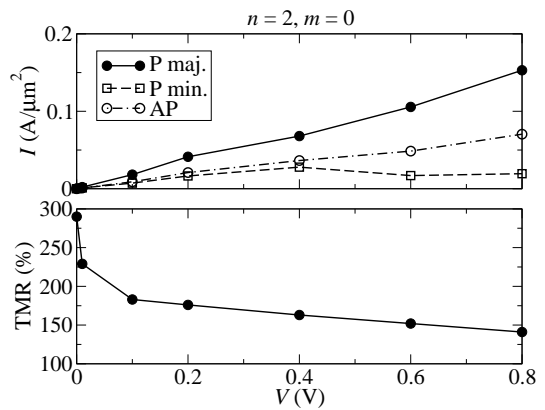}}
\caption{Current density and TMR as a function of bias voltage of the 
single-barrier junction with $n$=2 ML.}
\label{Fig2}
\end{figure}

As an example Figure~\ref{Fig2} shows the current-voltage curve for $n$=2 ML (4 \AA), together with the 
corresponding TMR. At zero bias, the TMR is calculated from the transmission probabilities, Eq.~(\ref{k}), evaluated at $E_\mathrm{F}$. It is seen that the currents for each spin channel 
are almost linear functions of $V$
over the bias range investigated. However, the P minority current saturates at $V=0.4$~Volt while the P majority 
and the AP currents keep increasing beyond that value. The competition between the rates of increase as a function of bias of the P majority and AP currents produces a rapid decay of the TMR and we calculate $V_{1/2}$ 
being at around 0.7~Volt.  
The TMR peaks at zero bias where it reaches up to 290 $\%$ but, because of its almost exponential decay, it
approaches the mean value over the bias range investigated of 160 $\%$ within only 
0.2~Volt. All of these features are in good agreement with experimental data \cite{yuasanat,tiusan06,noz05} and
early theoretical results \cite{mathon01,waldron}. It is important to note also that, at very low bias voltages, 
the P minority current is slightly larger than the AP current. As it was shown by Rungger {\it et al.} \cite{rungger}, 
this large P minority current near to zero-bias is due to the resonance of a Fe minority surface state through the 
MgO barrier and it is washed out as soon as a voltage is applied across the device.

\section{Results: Double-barrier tunnel junctions}

\begin{figure}[htb]
\epsfxsize=7.0cm
\centerline{\epsffile{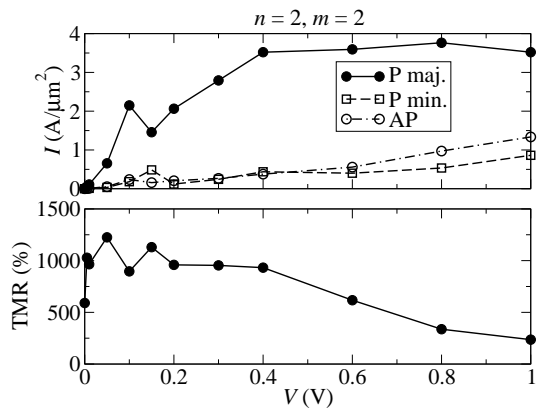}}
\caption{Current density and TMR as a function of bias voltage for a double-barrier magnetic tunnel junction with 
$n$=2 MLs and $m$=2 MLs.}
\label{Fig3}
\end{figure}

\begin{figure}[htb]
\epsfxsize=7.0cm
\centerline{\epsffile{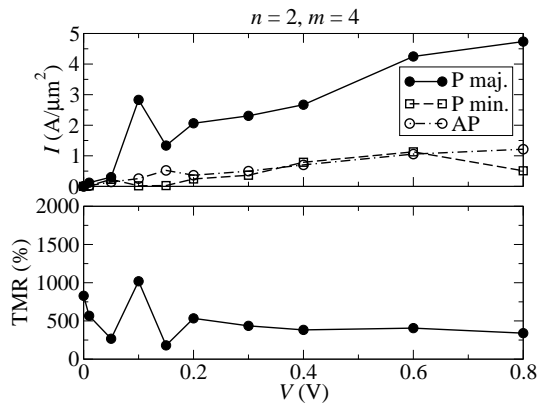}}
\caption{Current density and TMR as a function of bias voltage for a double-barrier magnetic tunnel junction with 
$n$=2 MLs and $m$=4 MLs.}
\label{Fig4}
\end{figure}

Figures \ref{Fig3} and \ref{Fig4} show the current density and the TMR for junctions with
$n$=2 MLs and $m$=2 MLs (2.87 \AA) and $m$=4 MLs (5.74 \AA), respectively. 
The first thing to notice is the appearance of oscillations in the TMR at low bias. These are not 
present in the SBMTJ, and stem from peaks in the low bias $I$-$V$ curve for all the spin channels. 
It is seen that the P majority current has a large peak at $V=0.1$~Volt, present for both values of $m$. 
Although the P minority and the AP currents also present peaks as a function of bias, these are considerably 
smaller than those in the P majority channel. A difference between the two junctions investigated is that
the bias at which the peaks appear depends on the number of monolayers $m$ of the in-between Fe layer.
Thus, the TMR oscillations as a function of bias are different in the two cases. 
In particular, for $m$=2 there are two TMR peaks at $V=0.05$~Volt and at $V=0.15$~Volt, at which the 
TMR reaches values close to 1250~$\%$. At $V=0.1$~Volt, the P majority current peak competes with that of
the AP configuration, resulting in the actual decrease of the TMR. In contrast, for $m$=4, there is a very 
large TMR peak at $V=0.1$~Volt, that clearly originates from the corresponding one in the P majority current. 

The peak in the P majority current of DBMTJs is suggestive of resonant tunneling through a QWS positioned 
in the in-between Fe slab. This is consistent with the experimental reports of Nozaki and coworkers \cite{nozosc} for 
Fe/MgO (001) DBMTJs with Fe islands as the in-between slab. They observed small conductance oscillations 
with bias {\it only} in the P configuration. In particular, for the thinnest in-between Fe slab investigated 
(1~nm), they observed two peaks in the P differential conductance at $V\sim$0.15~Volt and $V\sim$0.4~Volt.
Very recently, Wang {\it et al} \cite{wang} have theoretically shown that these conductance oscillations could be 
fit to majority QWSs of $\Delta_1$ symmetry at the $\Gamma$ point, once the level shift produced by 
charging effects is taken into account.  
As it can be seen in Fig. 2(a) of reference \cite{wang}, for ultrathin in-between Fe slabs 
there is a majority QWS at the $\Gamma$ point with an energy slightly less than $E_F$+0.1 eV. It is then tempting 
to ascribe our P majority peak at $V=$0.1~Volt to resonant tunneling through the above mentioned QWS.

\begin{figure}[htb]
\epsfxsize=8.0cm
\centerline{\epsffile{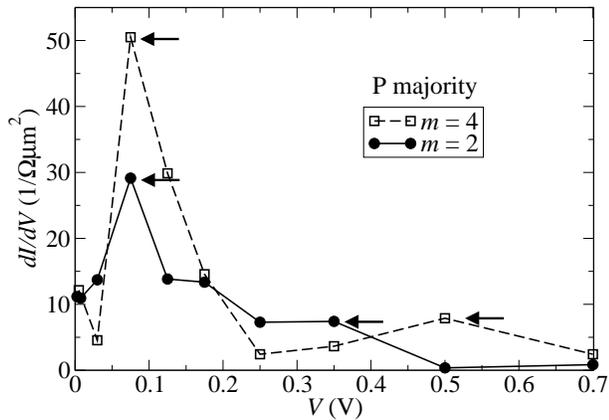}}
\caption{Differential conductance associated to the majority current in the P configuration for DBMTJs
with $m$=2 MLs, $m$=4 MLs, as a function of bias.}
\label{Fig5}
\end{figure}

Although our intention in this work is not to push for a quantitative agreement with the results of Nozaki 
{\it et al} \cite{nozosc}, it is interesting to compute the differential conductance of our DBMTJs, for the 
P majority channel. These are shown in figure \ref{Fig5}. The differential conductance is defined as 
$G=\mathrm{d}I/\mathrm{d}V$, and in this work it is simply calculated numerically from the $I$-$V$ curves. 
From Figure \ref{Fig5} it is seen that both DBMTJs show a conductance peak around $V=$~0.1~Volt. There 
is also a second conductance peak occurring at around $V=$~0.35~Volt for $m$=2 MLs and at around 
$V=$~0.5~Volt for $m$=4 MLs. Importantly the bias voltages at which these peaks occur are rather close to 
the experimental ones, even if our DBMTJs have significantly thinner in-between Fe slabs.

A second intriguing feature of the  DBMTJs is that the P majority current does not increase linearly with bias, 
in contrast to what happens for both the P minority and AP currents. As a consequence the TMR decay is not 
as severe as in the case of single-barrier junctions. In addition to that, it is important to report that the AP current for 
DBMTJs is significantly reduced with respect to that of the SBMTJs. For the simple SBMTJ investigated here, 
this is always larger than the P minority current, except at very low bias where the resonance of Fe minority surface state
plays the dominant role \cite{rungger}. In contrast, for double-barrier junctions the AP and the P minority currents
are almost equal to each other, even at relatively high voltages. This feature increases the bias 
voltage at which the TMR starts to decay pushing $V_{1/2}$ to higher biases. That is to say, our DBMTJs can 
{\it sustain} large TMR values even under the application of high bias 
voltages. As it was mentioned in the introduction, the reduction of the AP current is an indication 
of the SFE originated by the insertion of a magnetic slab. We then conclude that the spin-filter effect plays an 
important role in the TMR decay with bias in double-barrier junctions. 

By comparing the DBMTJs with the SBMTJ, it is clear that the TMR of DBMTJs does not only decay slower but 
it is also considerably higher than that of the SBMTJ with the same barrier thickness. This is due to the reduction of AP 
current in DBMTJs, originated from the spin-filter effect \cite{nos}. Therefore, the SFE in DBMTJs increases 
both $V_{1/2}$ and the TMR. These two results, together with the TMR oscillations at low bias mentioned 
above, are the most important features of our DBMTJs. It would be interesting to study the $I$-$V$ curves of 
DBMTJs with a very thick in-between Fe slab, since, according to our picture based on the SFE, the TMR decay with 
bias would be significantly reduced and the TMR values would increase. 

\section{Conclusions}

In summary, we have compared first principles calculations of the $I$-$V$ characteristics
of single and double Fe/MgO (001) magnetic tunnel junctions in the phase-coherent spin-conserving 
transport regime. Our results are in semi-quantitative agreement with recent experiments and 
demonstrate that: (i) the TMR for double-barrier junctions is significantly higher than that of 
single-barrier, not only at low but also for finite bias; (ii) the TMR decay with bias voltage 
is slower in double-barrier than in single-barrier junctions; (iii) double-barrier junctions 
exhibit low bias TMR oscillations. We have shown that the spin-filter effect is important in 
understanding the origin of the features (i) and (ii), while the third aspect is related to 
resonant electron tunneling through a majority spin QWS formed in the in-between Fe slab. 
In our opinion the first two aspects are the most relevant for real DBMTJs, since they are rather 
robust with respect to changes in the junction geometry. Thus, a possible route to obtain a very 
large TMR, that does not depend strongly on the bias voltage, is that of fabricating 
double-barrier junctions with very thick in-between magnetic slabs. This is a potentially
useful strategy for constructing spin-transfer element, where one needs to increase the
current density and preserve a high degree of spin-polarization.
 
This work was partially funded by UBACyT-X115, PIP-CONICET 6016, 
PICT 05-33304 and PME 06-117. I.~Rungger and S.~Sanvito thanks
Science Foundation of Ireland for financial support (grant SFI 07/IN.1/I945).
A. M. Llois belongs to CONICET (Argentina).

\end{document}